\documentclass[aps,nofootinbib,prd,superscriptaddress,tightenlines,notitlepage,twocolumn,showpacs]{revtex4-1}

\usepackage[colorlinks]{hyperref}
\usepackage{tabularx}
\usepackage{graphicx}
\usepackage{amssymb,bm,tensor}
\usepackage{xcolor}
\usepackage{amsmath}
\usepackage[varg]{txfonts}
\usepackage{enumerate}
\usepackage[normalem]{ulem}

\usepackage{scalerel,stackengine}
\stackMath
\newcommand\reallywidehat[1]{%
\savestack{\tmpbox}{\stretchto{%
  \scaleto{%
    \scalerel*[\widthof{\ensuremath{#1}}]{\kern-.6pt\bigwedge\kern-.6pt}%
    {\rule[-\textheight/2]{1ex}{\textheight}}
  }{\textheight}%
}{0.5ex}}%
\stackon[1pt]{#1}{\tmpbox}%
}

\newcommand{\AEI}{\affiliation{Max Planck Institute for Gravitational Physics
(Albert Einstein Institute), Am M\"uhlenberg 1, Potsdam D-14476, Germany}}

\newcommand{\UNLV}{\affiliation{Department of Physics and Astronomy, University
of Nevada, Las Vegas, NV 89154, USA}}

\newcommand{\KIAA}{\affiliation{Kavli Institute for Astronomy and Astrophysics,
Peking University, Beijing 100871, China}}

\newcommand{\DOA}{\affiliation{Department of Astronomy, School of Physics,
Peking University, Beijing 100871, China}}

\begin{document}

\title{A Bayesian Framework to Constrain the Photon Mass with a Catalog of Fast
Radio Bursts}
\author{Lijing Shao}\thanks{lijing.shao@aei.mpg.de}\AEI
\author{Bing Zhang}\thanks{zhang@physics.unlv.edu}\UNLV\KIAA\DOA

\begin{abstract}
  A hypothetical photon mass, $m_\gamma$, gives an energy-dependent light speed
  in a Lorentz-invariant theory. Such a modification causes an additional time
  delay between photons of different energies when they travel through a fixed
  distance.  Fast radio bursts (FRBs), with their short time duration and
  cosmological propagation distance, are excellent astrophysical objects to
  constrain $m_\gamma$. Here for the first time we develop a Bayesian framework
  to study this problem with a catalog of FRBs. Those FRBs with and without
  redshift measurement are both useful in this framework, and can be combined
  in a Bayesian way. A catalog of 21 FRBs (including 20 FRBs without redshift
  measurement, and one, FRB~121102, with a measured redshift $z=0.19273 \pm
  0.00008$) give a combined limit $m_\gamma \leq 8.7 \times 10^{-51}\, {\rm
  kg}$, or equivalently $m_\gamma \leq 4.9 \times 10^{-15}\, {\rm eV}/c^2$
  ($m_\gamma \leq 1.5\times10^{-50} \, {\rm kg}$, or equivalently $m_\gamma
  \leq 8.4 \times 10^{-15} \,{\rm eV}/c^2$) at 68\% (95\%) confidence level,
  which represents the best limit that comes purely from kinematics.  The
  framework proposed here will be valuable when FRBs are observed daily in the
  future.  Increment in the number of FRBs, and refinement in the knowledge
  about the electron distributions in the Milky Way, the host galaxies of FRBs,
  and the intergalactic median, will further tighten the constraint.
\end{abstract}

\maketitle


\section{Introduction}
\label{sec:intro}

The special relativity postulates the ``constancy of light speed'' as a
fundamental principle in physics~\cite{Einstein:1905ve}. It is extended into
the general relativity and quantum field theories. In quantum mechanics, the
particle-wave duality translates the constancy of light speed into the
``masslessness of photons''~\cite{Weinberg:1995mt}.  Nevertheless, there exist
theories with massive photons. The Maxwell-de Broglie-Proca theory is a famous
example where photons obtain mass in the cost of the gauge
invariance~\cite{Proca:1900nv, deBroglie:1922zz}. Another example where
``effectively massive photons'' arise due to a possible oscillation between the
canonical U(1) photons and hypothetical U$'$(1)
``photons''~\cite{Georgi:1983sy}.  Despite the celebrated success of the
postulate, these scenarios are interesting and worthy to investigate, whereas
the ultimate word on the photon mass roots in empirical facts.

Great efforts were put to various kinds of experiment to push the empirical
boundary on the ``masslessness of photons''~\cite{Goldhaber:2008xy,
Spavieri:2011zz}. These tests start from early experiment in testing the
$\propto r^{-2}$ Coulomb law~\cite{Williams:1971ms}, to Schumann resonance in
cavity~\cite{Kroll:1971wi}, gravitational deflection of electromagnetic
waves~\cite{Lowenthal:1973ka}, mechanical stability of magnetized gas in
galaxies~\cite{Chibisov:1976mm}, Jupiter's magnetic field~\cite{Davis:1975mn},
toroid Cavendish balance~\cite{Lakes:1998mi, Luo:2003rz}, magneto-hydrodynamics
of the solar wind~\cite{Ryutov:1997zz, Ryutov:2007zz}, black hole
bombs~\cite{Pani:2012vp}, and spindown of a white-dwarf
pulsar~\cite{Yang:2017ece}. These tests involve the anomalous {\it dynamics}
introduced by the mass term of photons, while there exist cleaner tests which
only involve the anomalous {\it kinematics} introduced by the
mass~\cite{Lovell:1964, Wu:2016brq, Bonetti:2016cpo, Bonetti:2017pym}, thus
independent of the underlying massive photon theory. The latter kind of tests
use the propagation of photons through a cosmological distance, and examine the
time delay between photons with different energies. In this paper we will study
the empirical mass constraint on the photon mass from the propagation of
electromagnetic waves of fast radio bursts (FRBs)~\cite{Lorimer:2007qn,
Thornton:2013iua, Champion:2015pmj}.

If photon has a mass $m_\gamma$, the Lorentz-invariant dispersion relation
reads,
\begin{eqnarray}\label{eq:dispersion}
  E^2 &=& p^2 c_0^2 + m_\gamma^2 c_0^4 \,,
\end{eqnarray}
where $c_0$ is the limiting velocity for high energy photons. The group
velocity for a photon with energy $E = h\nu$ is,
\begin{eqnarray}\label{eq:velocity}
  v &\equiv& \frac{\partial E}{\partial p} = \frac{pc_0}{E} c_0 \simeq c_0
  \left[ 1 - \frac{1}{2} \left( \frac{m_\gamma c_0^2}{h\nu} \right)^2 \right]
  \,,
\end{eqnarray}
where the last derivation holds when $m_\gamma \ll h\nu / c_0^2 \simeq
7\times10^{-42} \, {\rm kg} \, \left( \frac{\nu}{\rm GHz} \right)$. It is
evidently clear from Eq.~(\ref{eq:velocity}) that when the energy of a photon
is smaller, the relative modification is larger.\footnote{This is opposite to
  the test of Lorentz invariance violation with light
  propagation~\cite{AmelinoCamelia:1997gz, Mattingly:2005re, Shao:2009bv},
where high energy photons are preferred to put constraints.} Because we here
study the test that involves the accumulative time delay from light
propagation, it is easy to understand that (i) finer time structure of the
event, (ii) longer propagation distance, and (iii) lower energy for photons,
define the figure of merit of the test.  With this setting, FRBs provide a
superb celestial laboratory for testing the photon mass, because ---
\begin{enumerate}
  \item they are very short and even not temporally resolved by the receivers
    in general, and
  \item they are believed to be cosmological objects with non-negligible
    redshifts, and
  \item they are observed at radio frequency, which leverages the smallness of
    $m_\gamma$ in Eq.~(\ref{eq:velocity}).
\end{enumerate}

\begin{figure}
  \centering
  \includegraphics[width=8cm]{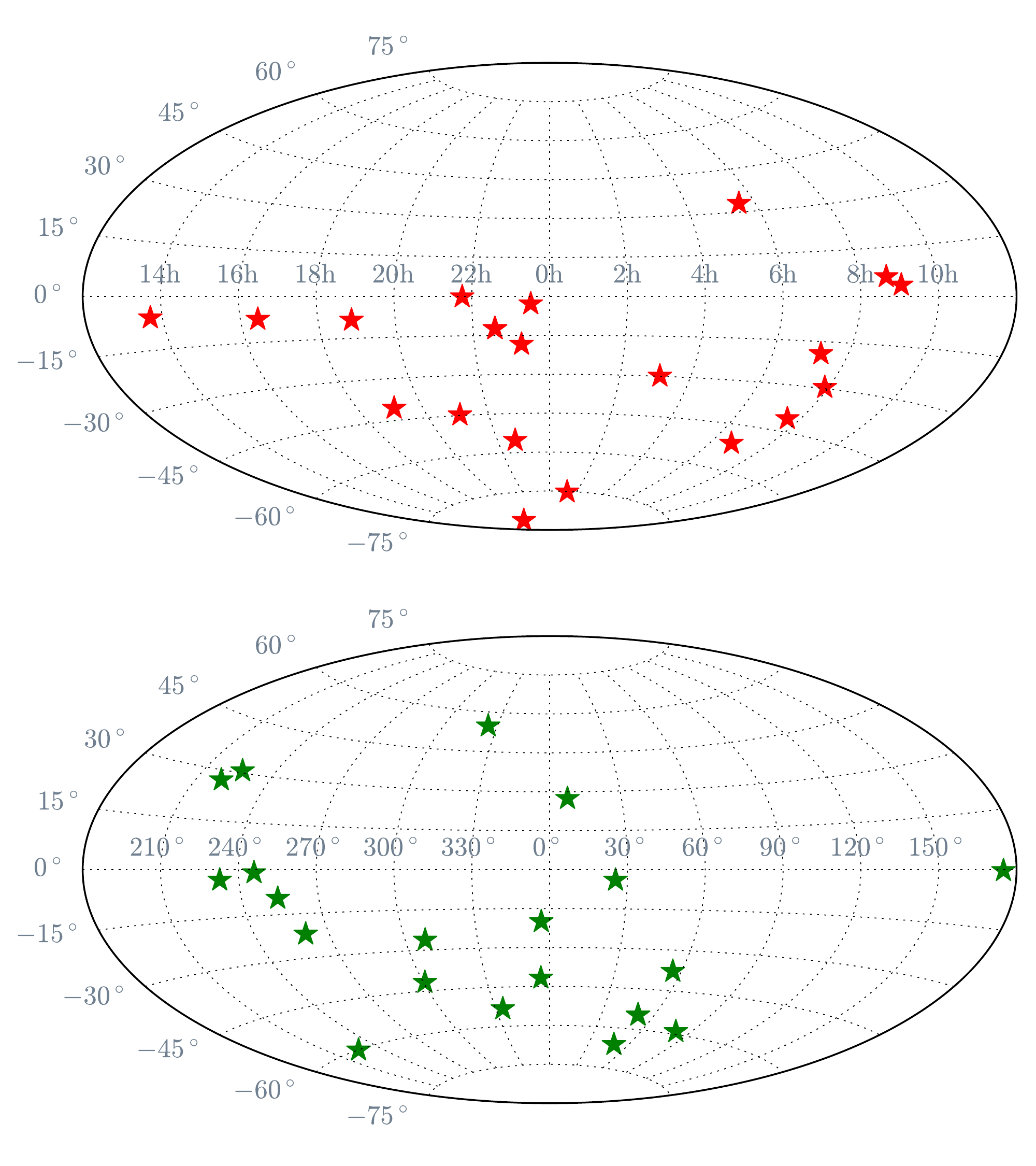}
  \caption{Distribution of FRBs that are used in constraining the photon
    mass~\cite{Petroff:2016tcr} in ({\it upper}) celestial coordinate, and
    ({\it lower}) galactic coordinate. \label{fig:frbsky}}
\end{figure}

The first work using FRBs to constrain the photon mass, performed by
\citet{Wu:2016brq} and \citet{Bonetti:2016cpo}, used
FRB~150418~\cite{Keane:2016yyk}. However, the measurement of the redshift for
this FRB was challenged with a flare in a radio-variable active galactic
nucleus~\cite{Williams:2016zys, Vedantham:2016ufj}, and now the measurement is
generally believed to be unreliable~\cite{Chatterjee:2017dqg}.  A reliable
measurement of the redshift was made for FRB~121102~\cite{Chatterjee:2017dqg,
Tendulkar:2017vuq}, and \citet{Bonetti:2017pym} followed up the measurement to
constrain the photon mass to be $m_\gamma \leq 3.9 \times 10^{-50} \, {\rm
kg}$, or equivalently, $m_\gamma \leq 2.2 \times 10^{-14} \, {\rm eV}/c^2$. The
method proposed in these papers needs a measurement of the redshift for FRBs,
however, up to now, only one FRB is fortunate enough to identify the host
galaxy and gets a redshift measurement. Since the localization of an FRB is
facilitated if the source is repeating and since none of the other FRBs are
observed to repeat so far, the sample of redshift-measured FRBs may not grow
significantly in the near future~\cite{Palaniswamy:2017aze}. Therefore, we here
extend these work to FRBs where the redshift is not available. We construct a
Bayesian formula to derive a combined constraint from a catalog of FRBs, where
uninformative prior is made to the redshift.  Figure~\ref{fig:frbsky} shows the
sky distribution of FRBs that are used in this work (see also
Table~\ref{tab:frbcat})~\cite{Petroff:2016tcr}.

The paper is organized as follows. In the next section, the theoretical
framework for data analysis is laid which includes a hypothesis for the
$\nu^{-2}$-behaved time delay, and a Bayesian framework to constrain
$m_\gamma$.  In Section~\ref{sec:res} we examine our uninformative treatment of
the redshift with FRB~121102, and present the constraint that comes from a
combination of a catalog of 21 FRBs where only one of them has a redshift
measurement.  Section~\ref{sec:dis} summarizes the work and discusses future
prospects in constraining the photon mass with FRBs.

\begin{table*}
\caption{A catalog of
  FRBs\footnote{\url{http://www.astronomy.swin.edu.au/pulsar/frbcat/}}~\cite{Petroff:2016tcr}
  that are used to constrain the photon mass. Sky position is given in right
  ascension, $\alpha$, and declination, $\delta$, at vernal equinox of J2000.0
  epoch. Dispersion measure is in unit of ${\rm pc\,cm}^{-3}$, where ${\rm
  DM}_{\rm obs}$ is from the fitting of the $\nu^{-2}$ behavior in the
  frequency-dependent time delay, and ${\rm DM}_{\rm NE2001}$ is based on the
  NE2001 Galactic electron density model~\cite{Cordes:2002wz}. $z_{\rm max}$ is
  the upper limit on the true redshift, obtained by assuming that the excess
  dispersion measure, ${\rm DM}_{\rm excess} \equiv {\rm DM}_{\rm obs} - {\rm
  DM}_{\rm NE2001}$, entirely comes from the IGM; since here we consistently
  use the parameter given in the main text and the full expression of $H_e(z)$,
  their values are close to, but larger than, that given in the
  catalog~\cite{Petroff:2016tcr}.  \label{tab:frbcat}}
\centering
\def\arraystretch{1.3}
\setlength{\tabcolsep}{0.6cm}
\begin{tabularx}{\textwidth}{llllrr}
\hline
FRB & Telescope & Sky position $\left(\alpha,\delta\right)_{\rm J2000}$ & ${\rm
DM}_{\rm obs}$ & ${\rm DM}_{\rm NE2001}$ & $z_{\rm max}$ \\
\hline
010125~\cite{Burke-Spolaor:2014rqa} & Parkes & $(19^{\rm h}06^{\rm m}53^{\rm s}, -40^\circ37'14'')$ & $790.3\pm0.3$ & $110$ & 0.77 \\
010621~\cite{Keane:2011} & Parkes & $(18^{\rm h}52^{\rm m}05^{\rm s}, -08^\circ29'35'')$ & $748\pm3$ & $523$ & 0.27 \\
010724~\cite{Lorimer:2007qn} & Parkes & $(01^{\rm h}18^{\rm m}06^{\rm s}, -75^\circ12'18'')$ & $375\pm3$ & $44.58$ & 0.38 \\
090625~\cite{Champion:2015pmj} & Parkes & $(03^{\rm h}07^{\rm m}47^{\rm s}, -29^\circ55'36'')$ & $899.55\pm 0.01$ & $31.69$ & 0.98 \\
110220~\cite{Thornton:2013iua} & Parkes & $(22^{\rm h}34^{\rm m}38^{\rm s}, -12^\circ23'45'')$ & $944.38\pm 0.05$ & $34.77$ & 1.02 \\
110523~\cite{Masui:2015kmb} & GBT & $(21^{\rm h}45^{\rm m}12^{\rm s}, -00^\circ09'37'')$ & $623.30\pm 0.06$ & $43.52$ & 0.66 \\
110626~\cite{Thornton:2013iua} & Parkes & $(21^{\rm h}03^{\rm m}43^{\rm s}, -44^\circ44'19'')$ & $723.0\pm 0.3$ & $47.46$ & 0.76 \\
110703~\cite{Thornton:2013iua} & Parkes & $(23^{\rm h}30^{\rm m}51^{\rm s}, -02^\circ52'24'')$ & $1103.6\pm 0.7$ & $32.33$ & 1.20 \\
120127~\cite{Thornton:2013iua} & Parkes & $(23^{\rm h}15^{\rm m}06^{\rm s}, -18^\circ25'38'')$ & $553.3\pm 0.3$ & $31.82$ & 0.60 \\
121002~\cite{Champion:2015pmj} & Parkes & $(18^{\rm h}14^{\rm m}47^{\rm s}, -85^\circ11'53'')$ & $1629.18\pm 0.02$ & $74.27$ & 1.77 \\
121102~\cite{Spitler:2014fla} & Arecibo, GBT & $(05^{\rm h}32^{\rm m}09^{\rm s}, +33^\circ05'13'')$ & $557\pm 2$ & $188$ & 0.43 \\
130626~\cite{Champion:2015pmj} & Parkes & $(16^{\rm h}27^{\rm m}06^{\rm s}, -07^\circ27'48'')$ & $952.4\pm 0.1$ & $66.87$ & 1.00 \\
130628~\cite{Champion:2015pmj} & Parkes & $(09^{\rm h}03^{\rm m}02^{\rm s}, +03^\circ26'16'')$ & $469.88\pm 0.01$ & $52.58$ & 0.48 \\
130729~\cite{Champion:2015pmj} & Parkes & $(13^{\rm h}41^{\rm m}21^{\rm s}, -05^\circ59'43'')$ & $861\pm 2$ & $31$ & 0.93 \\
131104~\cite{Ravi:2014mma} & Parkes & $(06^{\rm h}44^{\rm m}10^{\rm s}, -51^\circ16'40'')$ & $779\pm 3$ & $71.1$ & 0.80 \\
140514~\cite{Petroff:2014taa} & Parkes & $(22^{\rm h}34^{\rm m}06^{\rm s}, -12^\circ18'46'')$ & $562.7\pm 0.6$ & $34.9$ & 0.60 \\
150418~\cite{Keane:2016yyk} & Parkes & $(07^{\rm h}16^{\rm m}35^{\rm s}, -19^\circ00'40'')$ & $776.2\pm 0.5$ & $188.5$ & 0.67 \\
150807~\cite{Ravi:2016kfj} & Parkes & $(22^{\rm h}40^{\rm m}23^{\rm s}, -55^\circ16')$ & $266.5\pm 0.1$ & $70$ & 0.23 \\
160317~\cite{Caleb:2017vbk} & UTMOST & $(07^{\rm h}53^{\rm m}47^{\rm s}, -29^\circ36'31'')$ & $1165\pm 11$ & $319.6$ & 0.95 \\
160410~\cite{Caleb:2017vbk} & UTMOST & $(08^{\rm h}41^{\rm m}25^{\rm s}, +06^\circ05'05'')$ & $278\pm 3$ & $57.7$ & 0.26 \\
160608~\cite{Caleb:2017vbk} & UTMOST & $(07^{\rm h}36^{\rm m}42^{\rm s}, -40^\circ47'52'')$ & $682\pm 7$ & $238.3$ & 0.51 \\
\hline
\end{tabularx}
\end{table*}

\section{Theoretical Framework}
\label{sec:theory}

We review a hypothesis on the $\nu^{-2}$ behavior observed in the time delay
of FRBs in Section~\ref{sec:hypothesis}, and then construct a Bayesian
framework in Section~\ref{sec:bayes} to analyze the observed FRB data.

\subsection{A hypothesis on the time delay}
\label{sec:hypothesis}

\begin{figure}
  \centering
  \includegraphics[width=8cm]{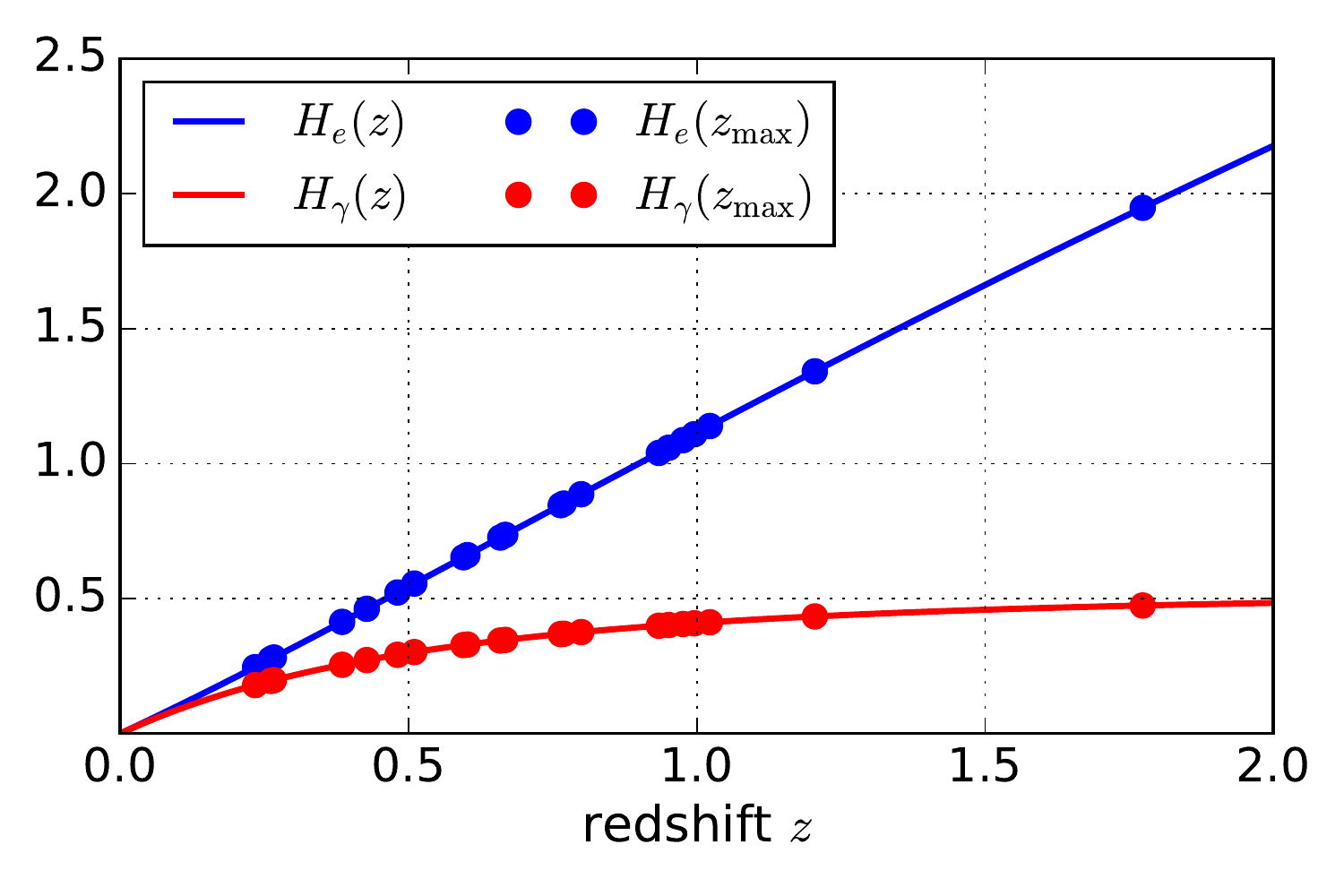}
  \caption{The dependence of functions $H_e(z)$ and $H_\gamma(z)$ on the
    redshift. Their function values evaluated at the maximum redshift, $z_{\rm
    max}$, for 21 FRBs in Table~\ref{tab:frbcat}, are shown with circles.
    \label{fig:he:hgamma}}
\end{figure}

Here we present the hypothesis of time delay for FRBs when photon has a mass,
$m_\gamma$~\cite{Wu:2016brq, Bonetti:2016cpo, Bonetti:2017pym}. The hypothesis,
${\cal H}$, will be used in the Bayesian inference in Section~\ref{sec:bayes}.

From observations, all FRBs show an indisputable $\nu^{-2}$-dependent time
delay, $\Delta t \propto \nu^{-2}$~\cite{Petroff:2016tcr, Katz:2016dti}. In our
hypothesis, we attribute the delay to two causes, (i) the propagation of
electromagnetic wave through ionized median, and (ii) the mass term of photon.
Some remarks come as follows ---
\begin{itemize}
    \item The interaction between the propagating electromagnetic wave and the
      ionized median introduces a time delay, $\Delta t_{\rm DM}$, for a photon
      with energy, $E=h\nu$, relative to a photon with an infinite
      energy~\cite{Lorimer:2012},
    \begin{eqnarray}\label{eq:deltaT:DM}
        \Delta t_{\rm DM} &=& \int \frac{{\rm d} l}{c_0} \frac{\nu_{\rm
        p}^2}{2\nu^2} \simeq 4.15\,{\rm ms} \left( \frac{\rm DM_{\rm
        astro}}{\rm pc\,cm^{-3}} \right) \left( \frac{\nu}{1\,{\rm GHz}}
        \right)^{-2} \,,
    \end{eqnarray}
    where the plasma frequency $\nu_{\rm p} \equiv \sqrt{n_e e^2 / 4\pi^2 m_e
    \epsilon_0}$ with $n_e$ the number density of electrons, $e$ the charge of
    an electron, $m_e$ the mass of an electron, and $\epsilon_0$ the
    permittivity of free space. The dispersion measure is defined as the
    integral of the electron number density along the path, ${\rm DM}_{\rm
    astro} \equiv \int n_e {\rm d}l$. In a cosmological setting, ${\rm DM}_{\rm
    astro} \equiv \int (1+z)^{-1} n_e^{\rm (0)} {\rm d}l$ where $z$ is the
    redshift and $n_e^{\rm (0)}$ is the electron number density in the rest
    frame~\cite{Deng:2013aga}.

    In Eq.~(\ref{eq:deltaT:DM}), different sources contribute to the dispersion
    measure, ${\rm DM}_{\rm astro}$~\cite{Petroff:2016tcr, Wu:2016brq,
    Bonetti:2016cpo, Bonetti:2017pym}, notably from the Milky Way, ${\rm
    DM}_{\rm MW}$, from the intergalactic median (IGM), ${\rm DM}_{\rm IGM}$,
    and from the host galaxy, ${\rm DM}_{\rm host}$. Therefore, the total
    dispersion measure reads,
    \begin{eqnarray}\label{eq:DM}
      {\rm DM}_{\rm astro} &=& {\rm DM}_{\rm MW} +{\rm DM}_{\rm IGM} + {\rm
      DM}_{\rm host} \,,
    \end{eqnarray}
    where we have included contributions from the host galaxy and the
    near-source plasma (e.g. supernova remnant, pulsar wind nebula, HII
    region~\cite{Yang:2017bls}) collectively in ${\rm DM}_{\rm host}$. We will
    present in Section~\ref{sec:bayes} how different pieces in the above
    equation are treated in a Bayesian framework.

  \item With Eq.~(\ref{eq:velocity}), it is straightforward to show that, in
    the $\Lambda$CDM universe, a photon with energy, $E = h \nu$, propagates
    slower relative to that with an infinite energy by~\cite{Jacob:2008bw,
    Wu:2016brq, Bonetti:2016cpo, Bonetti:2017pym},
    \begin{eqnarray}\label{eq:deltaT:gamma}
      \Delta t_{m_\gamma} &=& \frac{1}{2H_0} \left( \frac{m_\gamma c_0^2}{h
      \nu} \right)^2 H_\gamma(z) \,,
    \end{eqnarray}
    where $H_0 = 67.74 \pm 0.46 \, {\rm km\,s^{-1}\,Mpc}^{-1}$ is the Hubble
    constant~\cite{Ade:2015xua}, and $H_\gamma(z)$ is a dimensionless function
    of the source redshifit $z$ (see Figure~\ref{fig:he:hgamma}),
    \begin{eqnarray}
        H_\gamma(z) &\equiv& \int_0^z \frac{1}{\left( 1 + z' \right)^2}
        \frac{{\rm d} z'}{\sqrt{\Omega_\Lambda + \left( 1 + z' \right)^3
          \Omega_{\rm m}}} \,,
    \end{eqnarray}
    where the matter energy density $\Omega_{\rm m} = 0.3089 \pm 0.0062$, and
    the vacuum energy density $\Omega_\Lambda = 0.6911 \pm
    0.0062$~\cite{Ade:2015xua}. In deriving Eq.~(\ref{eq:deltaT:gamma}), we
    have assumed a flat universe with the curvature energy density $\Omega_k =
    0$, and the expansion of universe has been properly taken into
    account~\cite{Jacob:2008bw, Wu:2016brq}.
\end{itemize}

In our hypothesis ${\cal H}$, the total time delay is,
\begin{eqnarray}\label{eq:deltaT}
  \Delta t &=& \Delta t_{\rm DM} + \Delta t_{\rm m_\gamma} \,.
\end{eqnarray}
The two terms in the above equation both depend on the frequency in the same
way, therefore, it conforms to the observational fact that the total time delay
$\Delta t \propto \nu^{-2}$. The observational dispersion measure, ${\rm
DM}_{\rm obs}$, is obtained from the fit of the $\nu^{-2}$ behavior from the
total time delay. In our hypothesis, it equals to,
\begin{eqnarray}\label{eq:DM:obs}
  {\rm DM}_{\rm obs} &=& {\rm DM}_{\rm astro} + {\rm DM}_\gamma \,,
\end{eqnarray}
where ${\rm DM}_{\rm astro}$ is given in Eq.~(\ref{eq:DM}), and we have denoted
the ``effective dispersion measure'' caused by the non-vanishing photon mass
as,
\begin{eqnarray}
  {\rm DM}_\gamma &\equiv& \frac{4\pi^2 m_e \epsilon_0 c_0^5}{h^2 e^2}
  \frac{H_\gamma(z)}{H_0} m_\gamma^2 \,.
\end{eqnarray}

\subsection{A Bayesian framework}
\label{sec:bayes}

In Bayesian analysis, given data ${\cal D}$ and a hypothesis  ${\cal H}$ (here
Eq.~\ref{eq:DM:obs}), the posterior distribution of $m_\gamma^2$ can be
obtained by,
\begin{equation}
  P\left( m_\gamma^2 | {\cal D}, {\cal H}, {\cal I} \right) =
  \frac{P\left( {\cal D} | m_\gamma^2, {\cal H}, {\cal I} \right)
  P\left( m_\gamma^2 | {\cal H},{\cal I}\right)}{P\left({\cal D} | {\cal
  H}, {\cal I} \right)} \,, \label{eq:bayes}
\end{equation}
where ${\cal I}$ is all other relevant prior background knowledge. In the above
equation, given ${\cal H}$ and ${\cal I}$, $P\left( m_\gamma^2 | {\cal H},{\cal
I}\right)$ is the prior on $m_\gamma^2$, $P\left( {\cal D} | m_\gamma^2, {\cal
H}, {\cal I} \right) \equiv {\cal L}$ is the likelihood for the data, and
$P\left({\cal D} | {\cal H}, {\cal I} \right)$ is the model evidence.

We choose a uniform prior on $m_\gamma$ in the range $ m_\gamma \in \left[
10^{-69}, 10^{-42} \right] \, {\rm kg}$. The lower end is chosen because it
corresponds to the ultimate upper limit that in principle we can probe in one
observation due to the uncertainty principle of quantum mechanics and the
finite age of our universe~\cite{Barrow:1984hh}, while the upper end is chosen
because beyond which the approximation in Eq.~(\ref{eq:velocity}) breaks down.
Such a wide prior across 27 orders of magnitude in $m_\gamma$ reflects our
conservativeness.

To display the likelihood ${\cal L}$ that we adopt in our calculation, we first
investigate different contributions in Eq.~(\ref{eq:DM}) ---
\begin{itemize}
  \item ${\rm DM}_{\rm MW}$: albeit with uncertainties, there are electron
    distribution models for the Milky Way that incorporate different
    astrophysical observational results~\cite{Cordes:2002wz, Yao:2017}. We here
    use the NE2001 model. For different FRBs, ${\rm DM}_{\rm MW}$ is calculated
    from their line of sight. We assign a $20\%$ uncertainty to the value given
    by the NE2001 model to account for possible model inaccuracy.

    In principle, there is an additional contribution from the Galactic
      halo, which is not captured by the NE2001 model because pulsars in
      general do not probe this regime~\cite{Cordes:2002wz}. The halo
      contribution is not easy to model, but in our case it could already have
      been included  {\it effectively} in ${\rm DM}_{\rm
      host}$~\cite{Yang:2017bls}, which is obtained from the excess dispersion
      measure of FRBs, ${\rm DM}_{\rm excess} \equiv {\rm DM}_{\rm obs} - {\rm
      DM}_{\rm NE2001}$ (see below). The large uncertainty that we assign to
      ${\rm DM}_{\rm MW}$ could also account for (at least part of) this
    unknown contribution.

  \item ${\rm DM}_{\rm IGM}$: the dispersion measure due to the intergalactic
    medium (IGM) is given by~\cite{Deng:2013aga},
    \begin{eqnarray}
      {\rm DM}_{\rm IGM} &=& \frac{3c_0 H_0 \Omega_{\rm b} f_{\rm IGM} }{8\pi G
        m_{\rm p}} \int_0^z \frac{ g(z) \left( 1 + z' \right) {\rm d}
      z'}{\sqrt{\Omega_\Lambda + \left( 1 + z' \right)^3 \Omega_{\rm m}}} \,,
    \end{eqnarray}
    where $m_{\rm p}$ is the mass of a proton, $\Omega_{\rm b} = 0.0486 \pm
    0.0007$ is the baryonic matter energy density~\cite{Ade:2015xua}, $f_{\rm
    IGM} \simeq 0.83$ is its fraction to the IGM~\cite{Fukugita:1997bi}, and
    \begin{eqnarray}
        g(z) &\equiv& \frac{3}{4} y_1 \chi_{e, {\rm H}}(z) + \frac{1}{8} y_2
        \chi_{e, {\rm He}}(z) \,,
    \end{eqnarray}
    where $y_1 \sim 1$, $y_2 \sim 1$, and the ionized fractions of IGM
    $\chi_{e, {\rm H}}(z) \simeq 1$ for hydrogen at $z < 6$~\cite{Fan:2006dp}
    and $\chi_{e, {\rm He}}(z) \simeq 1$ for helium at $z <
    2$~\cite{McQuinn:2009}.  Therefore for $z < 2$, we have,
    \begin{eqnarray}
      {\rm DM}_{\rm IGM}  &\simeq& \frac{3c_0 H_0 \Omega_{e,{\rm
      IGM}}}{8\pi G m_{\rm p}} H_e(z) \,,
      \label{eq:DM:IGM}
    \end{eqnarray}
    where the effective energy density of ionized baryons $\Omega_{e,{\rm IGM}}
    \simeq 0.035$~\cite{Ioka:2003fr, Inoue:2003ga, Deng:2013aga}.  To be
    conservative, we associate a $20\%$ uncertainty to $\Omega_{e,{\rm
    IGM}}$~\cite{Bonetti:2016cpo, Yang:2017bls}, in the hope that such a large
    uncertainty could account for, at least partially, the inhomogeneity of IGM
    along different line of sight. In Eq.~(\ref{eq:DM:IGM}), the dimensionless
    redshift function, $H_e(z)$, reads,
    \begin{eqnarray}
      H_e(z) &\equiv& \int_0^z \frac{ \left( 1 + z' \right) {\rm d}
    z'}{\sqrt{\Omega_\Lambda + \left( 1 + z' \right)^3 \Omega_{\rm m}}} \,.
    \end{eqnarray}
    In Figure~\ref{fig:he:hgamma} we depict $H_e(z)$ and $H_\gamma(z)$ with
    cosmological parameters from the $\Lambda$CDM model~\cite{Ade:2015xua}.
    Worthy to mention that, \citet{Bonetti:2016cpo, Bonetti:2017pym} pointed
    out that the different behavior of these two redshift functions might be
    able to break parameter degeneracy in testing the photon mass at the point
    when a handful measurements of redshift for FRBs become available. For now
    we leave this point to future work.

    To predict the contribution of ${\rm DM}_{\rm IGM}$, we need a redshift
    measurement, which is only available for FRB~121102 up to
    now~\cite{Chatterjee:2017dqg}.\footnote{We will not consider the redshift
      measurement for FRB~150418~\cite{Keane:2016yyk}, which is challenged with
      a flare in an active galactic nucleus~\cite{Williams:2016zys,
      Vedantham:2016ufj, Chatterjee:2017dqg}.  Nevertheless, the inclusion of
      this measurement into our framework is straightforward if the redshift
    measurement is proven genuine.} Due to our lack of knowledge, we assume a
    prior for $z$ such that the prior for the FRB's spatial distribution is
    uniform in the comoving spherical volume,
    \begin{eqnarray}
      V_{\rm c}^{\rm max} \equiv \left. \frac{4\pi}{3} D_{\rm c}^3 (z)
      \right|_{z = z_{\rm max}} \,,
    \end{eqnarray}
    where $D_c(z)$ is the comoving distance to a source at redshift $z$,
    \begin{eqnarray}
      D_{\rm c}(z) \equiv \frac{c_0}{H_0}
    \int_0^z \frac{ {\rm d}z'}{\sqrt{\Omega_\Lambda + \left( 1+z' \right)^3
      \Omega_{\rm m}}} \,,
    \end{eqnarray}
    and $z_{\rm max}$ is the maximum possible redshift value of redshift (see
    Table~\ref{tab:frbcat}) by setting ${\rm DM}_{\rm host} = 0 $ and $m_\gamma
    = 0$ in Eq.~(\ref{eq:DM:obs})~\cite{Petroff:2016tcr}.  Such a prior for $z$
    will be denoted as uniform for $V_{\rm c}(z) \in [0, V_{\rm c}^{\rm max}]$
    where $V_{\rm c}(z) \equiv \frac{4\pi}{3} D_{\rm c}^3(z)$ is the comoving
    spherical volume within redshift $z$. In Section~\ref{sec:dis} we will in
    addition present the results with a prior of $z$ that traces the star
    formation rate~\cite{Yuksel:2008cu}, and confirm the robustness of results.
    For FRB~121102, the measured redshift will be used as the prior for $z$
    (see Section~\ref{sec:121102}).

  \item  ${\rm DM}_{\rm host}$: From a ${\rm DM}_{\rm excess}$--$F_\nu$
    relation, \citet{Yang:2017bls} derived a statistical result, $\left\langle
    {\rm DM}_{\rm host} \right \rangle = 267^{+173}_{-111} \, {\rm
    pc\,cm}^{-3}$ in the rest frame of FRB, under the assumption that the
    isotropic-equivalent luminosity of FRBs has a characteristic value.
    Simulations show that a $\sim30\%$ Gaussian dispersion in $ {\rm DM}_{\rm
    host}$ still keeps the result valid~\cite{Yang:2017bls}. The conclusion in
    \citet{Yang:2017bls} is supported by the large scattering time of FRBs and
    the inferred $ {\rm DM}_{\rm host}$ from
    FRB~121102~\cite{Chatterjee:2017dqg, Yang:2017bls}. We will use this
    result in our estimation of ${\rm DM}_{\rm host}$, after multiplying it by
    a factor, $\left( 1 + z \right)^{-1}$, which takes the cosmological
    evolution into account.
    \footnote{Strictly speaking, the ${\rm DM}_{\rm host}$ obtained in
    \citet{Yang:2017bls} uses the assumption $m_\gamma=0$. A global re-analysis
    that closely follows the MCMC simulations in that work but allowing a
    nonzero $m_\gamma$ would be ideal, however, this goes beyond the scope of
    current work. Instead we perform the following simulation to assess the
    influence to our result from using the ${\rm DM}_{\rm host}$ in
    \citet{Yang:2017bls}. Notice that, when ${\rm DM}_{\rm host}$ is
    underestimated, the constraint on $m_\gamma$ is more conservative.
    Therefore, we perform the most conservative simulation that {\it
    artificially} sets ${\rm DM}_{\rm host} = 0$. We observe that, even under
    such an assumption, our results only change by a factor less than three.
    Consequently, the results in the paper are robust to possible changes in
    the ${\rm DM}_{\rm host}$ value we adopt.}
\end{itemize}

Finally the logarithm of likelihood is constructed as,
\begin{eqnarray}\label{eq:lnL}
  \ln {\cal L} &=& - \frac{1}{2} \sum_i \frac{\left( {\rm DM}_{\rm obs}^i -
  {\rm DM}_{\rm astro}^i - {\rm DM}_\gamma^i \right)^2}{\sigma_i^2}
  \,,
\end{eqnarray}
where $i$ indexes FRBs, ${\rm DM}_{\rm astro}$ is the dispersion measure
obtained with the above listed prescriptions in Markov-chain Monte Carlo (MCMC)
runs, $\sigma$ includes all uncertainties added in quadratic (including
uncertainties in ${\rm DM}_{\rm obs}$, ${\rm DM}_{\rm MW}$, ${\rm DM}_{\rm
IGM}$, and ${\rm DM}_{\rm host}$), and ${\rm
DM}_\gamma$ denotes the second term in Eq.~(\ref{eq:DM:obs}). In writing
Eq.~(\ref{eq:lnL}), an assumption is made that the observations of different
FRBs are independent.

\section{Results}
\label{sec:res}

\begin{figure*}
  \centering
  \includegraphics[width=15cm]{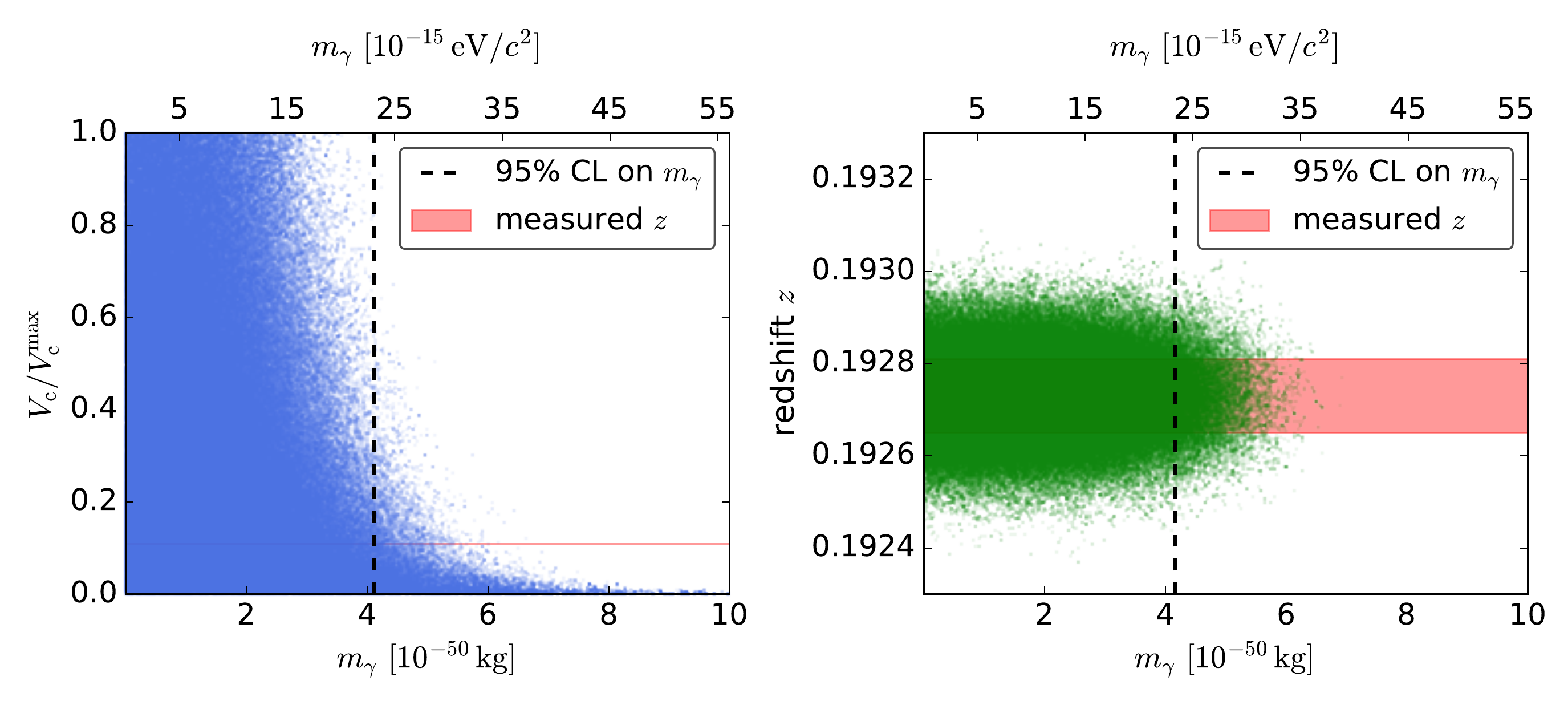}
  \caption{Distribution of MCMC samples from FRB~121102 ({\it left}) in the
  $m_\gamma$-$\left(V_{\rm c}/V_{\rm c}^{\rm max}\right)$ plane: without using
  the redshift measurement, and ({\it right}) in the $m_\gamma$-$z$ plane using
  the redshift measurement as the prior on $z$~\cite{Chatterjee:2017dqg}. The
  pink bands show the redshift measurement and its 1-$\sigma$ uncertainty
  obtained by~\citet{Chatterjee:2017dqg}; the band of $V_{\rm c}(z)/V_{\rm
  c}^{\rm max}$ is too narrow to be visible in the left panel.  The $95\%$ CL
  constraints on $m_\gamma$ are depicted as dashed lines.
  \label{fig:121102:post}}
\end{figure*}

\begin{figure}
  \centering
  \includegraphics[width=8cm]{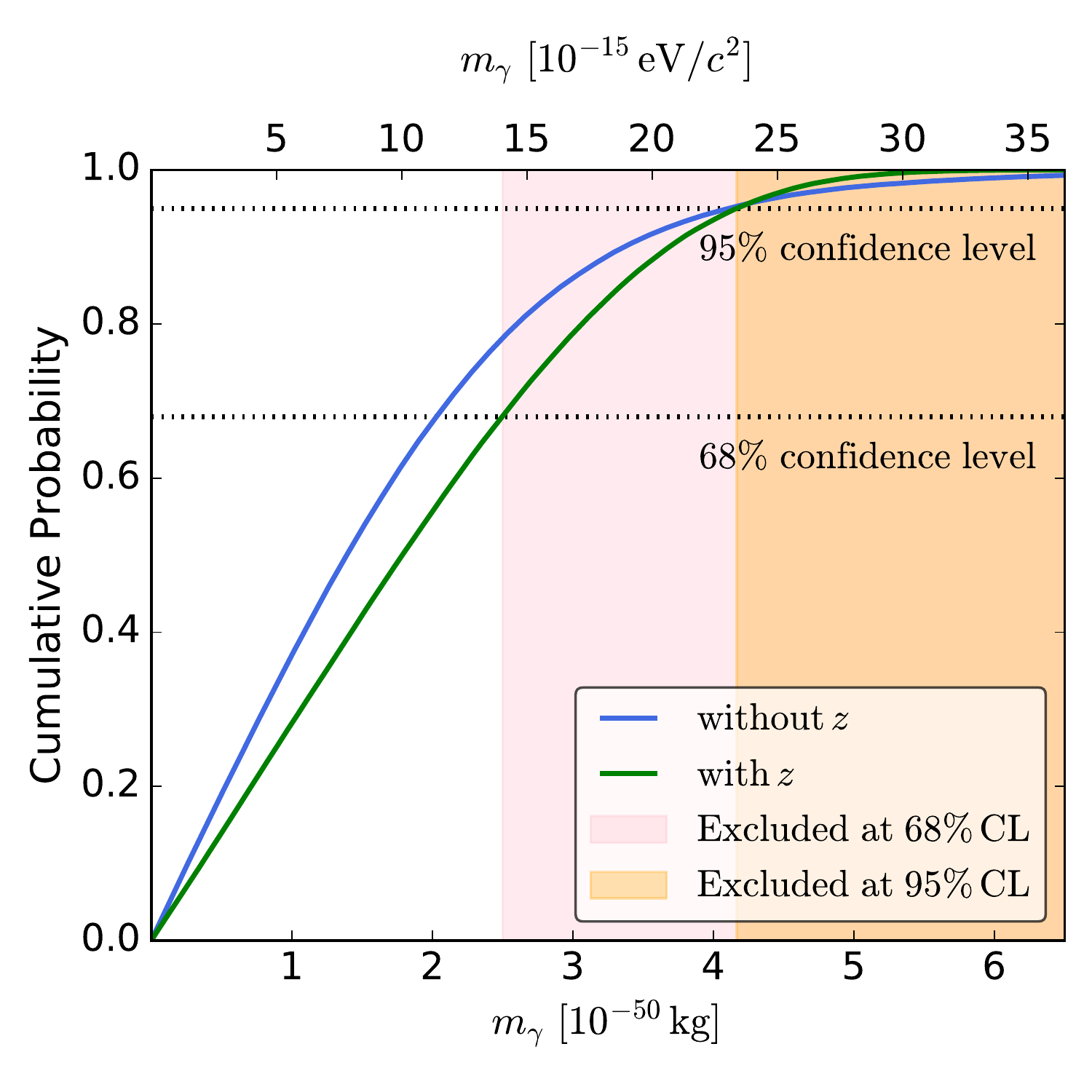}
  \caption{The cumulative posterior probability distributions on $m_\gamma$
    from FRB~121102 without using the redshift measurement (in blue) and using
    the redshift measurement (in green). The excluded values for $m_\gamma$ at
    $68\%$ and $95\%$ CLs are shown with shadowed areas for the case where the
    redshift is used.
    \label{fig:121102:cumProb}}
\end{figure}

\begin{figure*}
  \centering
  \includegraphics[width=15cm]{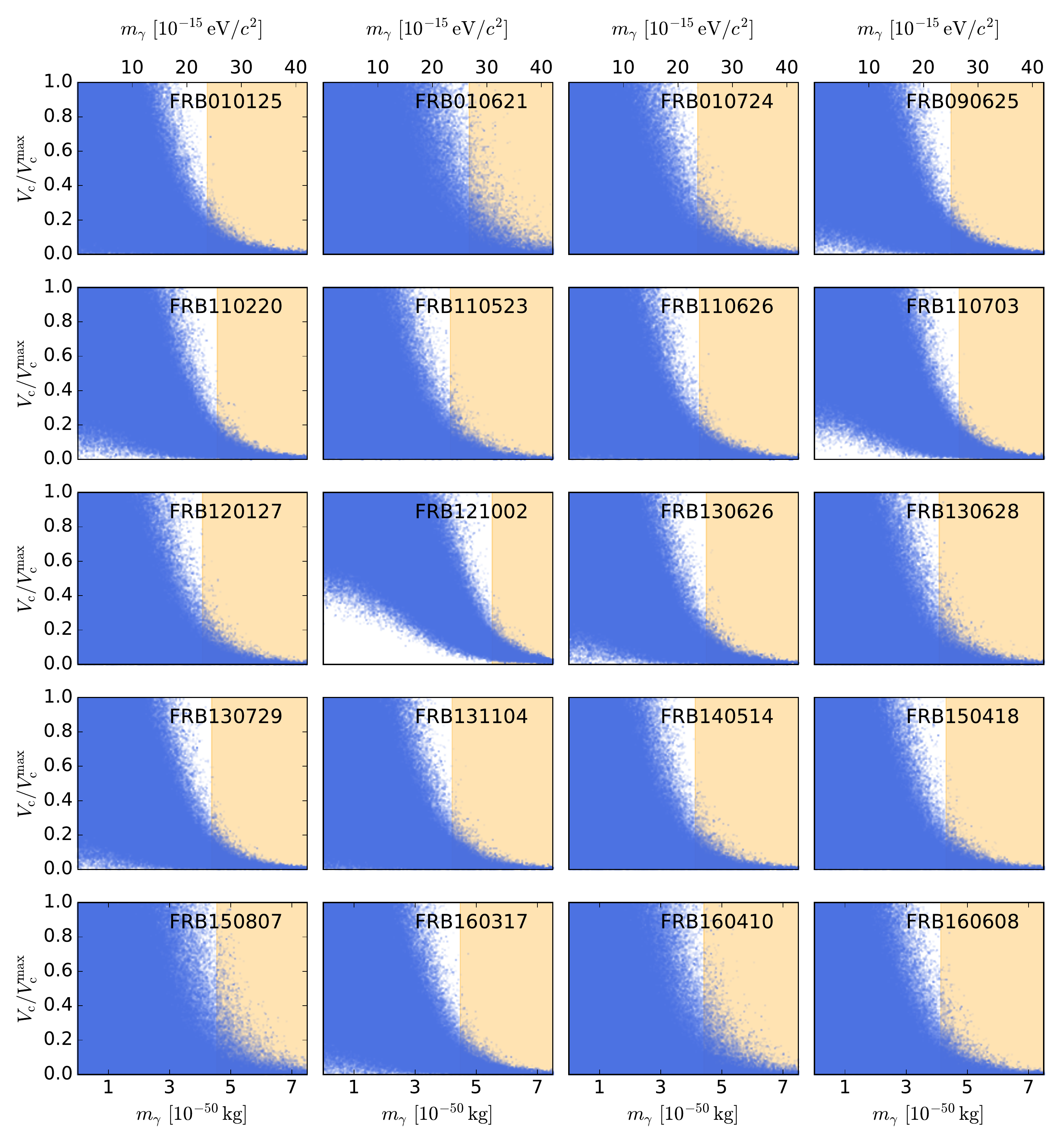}
  \caption{The MCMC samples in the $m_\gamma$-$\left(V_{\rm c}/V_{\rm c}^{\rm
  max}\right)$ plane for a catalog of FRBs in Table~\ref{tab:frbcat} except
  FRB~121102. The excluded regions at $95\%$ CL are shadowed for each FRB.
  \label{fig:frbs:post}}
\end{figure*}

\begin{figure}
  \centering
  \includegraphics[width=8cm]{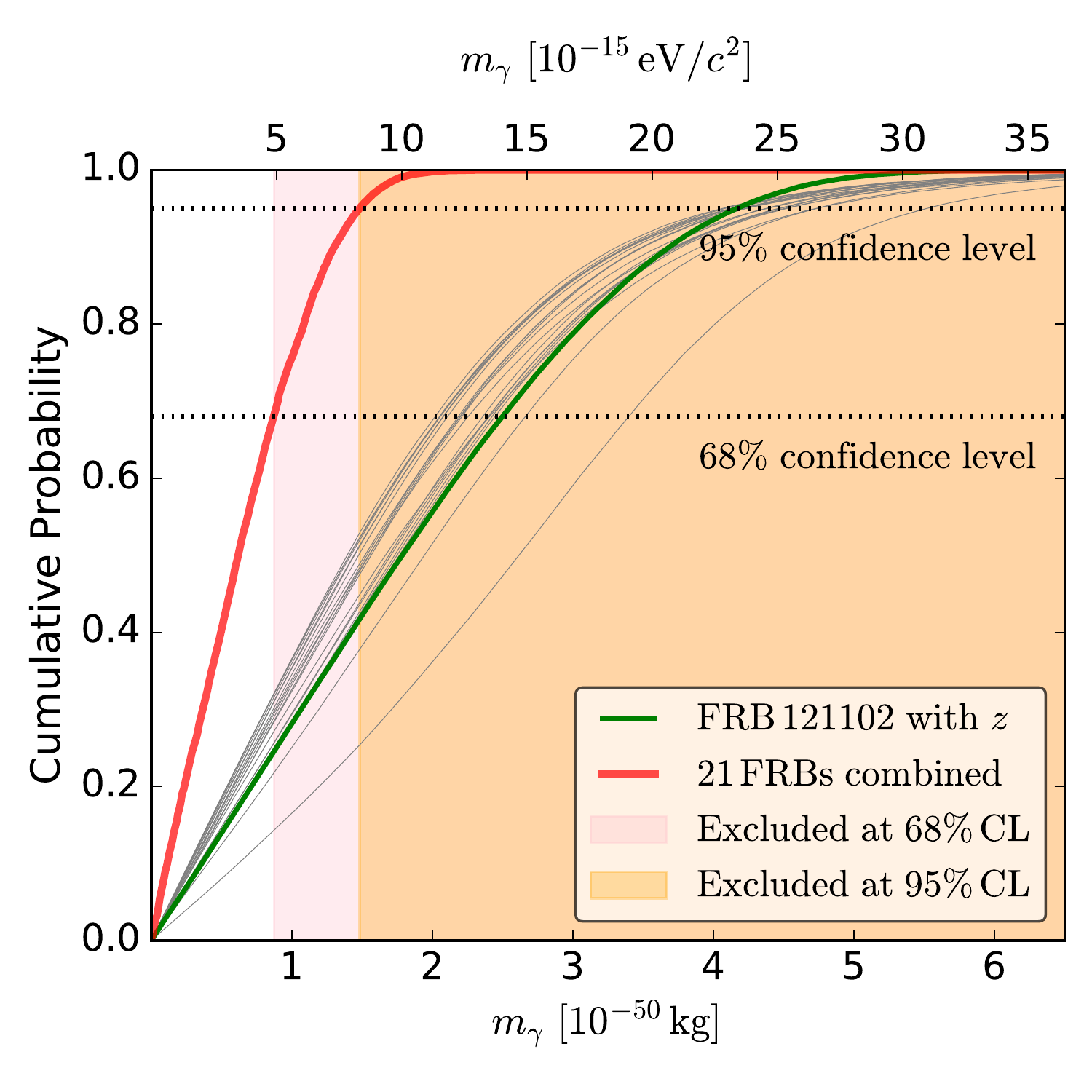}
  \caption{The cumulative posterior probability distributions for 20 FRBs (the
    catalog of FRBs in Table~\ref{tab:frbcat} except FRB~121102) are shown in
    grey. The same quantity is shown in green for FRB~121102 (the same green
    curve in Figure~\ref{fig:121102:cumProb}). The combination of these 21 FRBs
    is given in red. The excluded values for $m_\gamma$ at $68\%$ and $95\%$
    CLs are shown with shadowed areas for the combination.
    \label{fig:frbs:cumProb}}
\end{figure}

As said, we use MCMC techniques to explore the posterior in
Eq.~(\ref{eq:bayes}). Ideally, one would use the log-likelihood in
Eq.~(\ref{eq:lnL}) to simultaneously analyze all FRBs in one go, whereas here
the computational cost would be very high due to the large dimensionality of
the parameter space.  The dimensionality equals to the number of FRBs (their
redshifts) plus one (the photon mass $m_\gamma$).  We adopt a sub-optimal
strategy where the posteriors of $m_\gamma$, from different individual FRBs,
are combined {\it after} independent MCMC run is performed on each single
FRB~\cite{Lyons:2017eui}.  This is not a down-graded choice because we know
that the redshifts of different FRBs are unlikely to correlate with each other.
Such an approach is also the strategy adopted in constraining the strong
equivalence principle in Ref.~\cite{Stairs:2005hu}, the local Lorentz
invariance of gravity in Refs.~\cite{Shao:2013wga, Shao:2014bfa}, and the
parameterized tests of general relativity with the Advanced LIGO events in
Ref.~\cite{TheLIGOScientific:2016pea}.

We use the {\sc PYTHON} implementation of an affine-invariant MCMC ensemble
sampler~\cite{Goodman:2010, ForemanMackey:2012ig}, {\sc
emcee},\footnote{\url{http://dan.iel.fm/emcee}} to explore the posterior
distributions. This algorithm generally has better performance over the
traditional MCMC sampling methods (e.g., the Metropolis-Hasting algorithm), as
measured by the smaller autocorrelation time and fewer hand-tuning
parameters~\cite{ForemanMackey:2012ig}.  We set up MCMC runs to investigate the
$(m_\gamma, z)$ pair for each FRB. As mentioned, the priors are uniform in
$V_{\rm c}(z) \in [0, V_{\rm c}^{\rm max}]$ and in $m_\gamma \in
\left[10^{-69}, 10^{-42}\right] \, {\rm kg}$.  Each MCMC run samples the
posterior distribution according to Eq.~(\ref{eq:bayes}), with the
log-likelihood given by Eq.~(\ref{eq:lnL}), with 20 chains. For each FRB,
$10^6$ samples are accumulated. The first half of the samples are discarded as
the {\sc burn-in} phase~\cite{Brooks:2011}. We check the convergence of
different chains with the Gelman-Rubin statistic~\cite{Gelman:1992zz},
\begin{eqnarray}
  \hat R &\equiv& \sqrt{ \frac{\reallywidehat{{\rm Var}}\left( \bm{\Theta} |
  {\cal D} \right)}{W}} \,,
\end{eqnarray}
where the estimate of the marginal posterior variance for each parameter
$\bm{\Theta} \equiv \left\{ \theta_{ij} \right\}$ (indices $i,j$ denote the
$i$-th posterior sample in the $j$-th chain) is,
\begin{eqnarray}
  \reallywidehat{{\rm Var}}\left( \bm{\Theta} | {\cal D} \right) &\equiv&
  \frac{N-1}{N} W + \frac{1}{N} B \,,
\end{eqnarray}
with the between-chain variance, $B$, and the within-chain variance, $W$,
\begin{eqnarray}
  B &\equiv& \frac{N}{M-1} \sum_{j=1}^M \left( \frac{1}{N} \sum_{i=1}^N
  \theta_{ij} - \frac{1}{N} \sum_{i=1}^N \frac{1}{M} \sum_{k=1}^M \theta_{ik}
  \right)^2 \,, \\
  W &\equiv& \frac{1}{M} \sum_{j=1}^M \left[ \frac{1}{N-1} \sum_{i=1}^N \left(
    \theta_{ij} - \frac{1}{N} \sum_{k=1}^N \theta_{kj}
   \right)^2 \right] \,,
\end{eqnarray}
where $M$ and $N$ are respectively the number of chains (in our case $M=20$)
and the number of samples per chain (in our case $N=25000$ after discarding the
{\sc burn-in} samples). Our convergence test shows $\hat R \lesssim 1.002$ for
all cases, indicating very good convergence in MCMC runs.  The posteriors and
constraints on $m_\gamma$ are presented in the following subsections.

\subsection{Limit from FRB~121102}
\label{sec:121102}

Because the redshift of FRB~121102 was measured to great precision
in~Refs.~\cite{Chatterjee:2017dqg, Tendulkar:2017vuq}, $z=0.19273 \pm 0.00008$,
we would like to compare the constraints on the photon mass with and without
this measurement.  By including the redshift measurement, we mean using a
Gaussian prior for $z$, centered around its measured value with a spread of the
uncertainty.

In Figure~\ref{fig:121102:post} we show the samples returned by the MCMC
sampler (after discarding the {\sc burn-in} samples) in both cases for
FRB~121102. We immediately see that if the samples are marginalized over the
photon mass, priors on $z$ are more or less recovered in both cases. This means
that the Bayesian framework proposed here does not add more information to the
redshift, as it should not.

In Figure~\ref{fig:121102:cumProb} we show the accumulative posterior
probability on $m_\gamma$, marginalized over the redshift $z$, for both cases.
As we can see, the result from the use of redshift measurement is very close to
the one that does not use the redshift. We read out, at a 95\% confidence
level,
\begin{eqnarray}\label{eq:121102:woz}
  m_\gamma &\leq& 4.1 \times 10^{-50} \, {\rm kg} \,,
\end{eqnarray}
when using an uninformative uniform prior on $V_{\rm c}(z)$, and
\begin{eqnarray}\label{eq:121102:wz}
  m_\gamma &\leq& 4.2 \times 10^{-50} \, {\rm kg} \,,
\end{eqnarray}
when $z=0.19273 \pm 0.00008$ is used. The latter agrees well with the result
presented by \citet{Bonetti:2017pym} for this FRB with a less sophisticated
method. The marginalized 1\,D probability distribution on $m_\gamma$ with the
uninformative prior has a long tail which reflects our ignorance in the
redshift. The ultimate closeness of the results in
Eqs.~(\ref{eq:121102:woz}--\ref{eq:121102:wz}) is  a bit coincident, but it
also shows the reasonableness of the use of the  uninformative prior.

\subsection{Limits from individual FRBs}

Except for FRB~121102 discussed above, the other 20 FRBs in the catalog (see
Table~\ref{tab:frbcat}) unfortunately have no redshift
measurement~\cite{Petroff:2016tcr}. Therefore, we can only rely on the
uninformative priors. The distribution of MCMC samples are shown in
Figure~\ref{fig:frbs:post} for these FRBs. As in the FRB~121102 case with the
uninformative prior, the distributions have long tails towards large
$m_\gamma$.  Especially for the FRBs with large $z_{\rm max}$ (e.g., FRB~110703
and FRB~121002), a large $m_\gamma$ is needed to account for part of the
dispersion measure in ${\rm DM}_{\rm obs} - {\rm DM}_{\rm MW}$ when the
redshift is very small, as expected. From their panels in
Figure~\ref{fig:frbs:post}, we see that some regions with small $z$ and small
$m_\gamma$ have no support from MCMC runs. Because of the conservatively large
uncertainties that we use in ${\rm DM}_{\rm MW}$ and in ${\rm DM}_{\rm IGM}$,
these FRBs individually only constrain $m_\gamma$ at $\sim 5 \times
10^{-50}\,{\rm kg} $ at 95\% confidence level, as shown by the shadowed regions
in the figure.  The uncertainties in these two terms dominate the test, hence,
in terms of order of magnitude, all FRBs here have comparable constraints.

\subsection{Combined limit from a catalog of FRBs}

We now have 21 individual constraints on $m_\gamma$. Assuming that these FRBs
are independent, we can combine their posterior distributions, in the spirit of
Eq.~(\ref{eq:bayes}).  Similar combination of posteriors was done in
Refs.~\cite{Stairs:2005hu, Shao:2013wga, Shao:2014bfa,
TheLIGOScientific:2016pea} under different subjects.  Here, since for
FRB~121102 a reliable redshift is available~\cite{Chatterjee:2017dqg,
Tendulkar:2017vuq}, we use the result that takes this measurement into account.
In Figure~\ref{fig:frbs:cumProb}, we plot the marginalized accumulative
posterior distributions for 20 FRBs (see Figure~\ref{fig:frbs:post}) in gray,
that generally give $m_\gamma \lesssim 5 \times 10^{-50}\, {\rm kg}$ at 95\%
confidence level, and the accumulative posterior distribution for FRB~121102 in
green (same as the green curve in Figure~\ref{fig:121102:cumProb}), that gives
$m_\gamma \leq 4.2 \times 10^{-50} \, {\rm kg}$ at 95\% confidence level (see
Eq.~\ref{eq:121102:wz}).  We also give the accumulative posterior distribution
that combines the 21 FRBs with a red curve in the figure.  In the Bayesian
sense, it is unlikely that multiple FRBs reside in the long tails of their
distributions.  This result demonstrates the collective power of these
``deceptively boring'' FRBs that have no redshift measurement. It has strong
implication for future study using FRBs to constrain the photon mass.  The
final combination of 21 FRBs  (the red curve in Figure~\ref{fig:frbs:cumProb})
give a tight constraint on $m_\gamma$,
\begin{eqnarray}
  m_\gamma \leq 8.7 \times 10^{-51} \, {\rm kg} \simeq 4.9\times10^{-15} \,
  {\rm eV}/c^2 \,, \label{eq:limit:68} \\
   m_\gamma \leq 1.5 \times 10^{-50} \, {\rm kg} \simeq 8.4 \times 10^{-15} \,
   {\rm eV}/c^2  \label{eq:limit:95} \,,
\end{eqnarray}
at 68\% and 95\% confidence levels respectively.  These limits improve over
previous results that only used a single FRB~\cite{Wu:2016brq, Bonetti:2016cpo,
Bonetti:2017pym} by a factor of $\sim4$.

\section{Discussions}
\label{sec:dis}

Recently, the first direct observation of gravitational waves from a binary
black hole merger at a cosmological distance by the Advanced LIGO puts a
constraint on the graviton mass, $m_g \leq 1.2 \times 10^{-22} \, {\rm
eV}/c^2$, at 90\% confidence level~\cite{TheLIGOScientific:2016src}. Because
most of the power of the gravitational-wave event is at ${\cal O}(10^2) \, {\rm
Hz}$, even significantly lower than the ${\cal O}(10^9) \, {\rm Hz}$ radio
waves we here use to constrain the photon mass, a tighter constraint on the
mass is expected (cf. discussion below Eq.~\ref{eq:velocity}). Nevertheless,
the constraints on the photon mass (see Eqs.
\ref{eq:limit:68}--\ref{eq:limit:95}) pertain to a different sector of species.
As far as we are aware, this is the tightest limit on the photon mass that is
obtained solely depending on the propagation kinematics, therefore completely
avoiding assumptions about the underlying dynamical theory for the massive
photon.

The Bayesian framework proposed here will be even more valuable in future, when
more and more FRB observations become available (for example, with the ALERT
project\footnote{\url{http://www.alert.eu.org/}}). The improvement could come
from ---
\begin{enumerate}
  \item new discovery of more FRBs;
  \item more coincident measurements of FRB redshift;
  \item a better understanding of the various astrophysical contributions to
    the observed dispersion measure, including those from the Milky Way, the
    IGM, and the host galaxies of FRBs.
\end{enumerate}
If future FRBs are observed with the same quality as current ones, we expect a
rough ${\cal N}^{-1/2}$ improvement on the photon mass where ${\cal N}$ is the
number of FRBs. The improvement from the measurement of the redshift is very
hard to predict. It depends on the redshift value that is measured. For
example, in the case of FRB~121102, the measured redshift $z \simeq 0.19$
resides in the lower end of its possible values up to $z_{\rm max} \simeq
0.43$.  Taking the uniform prior in volume into account, one would expect a
chance of approximately $0.19^3 / 0.43^3 \simeq 9\%$ to observe a redshift as
low as $0.19$ for FRB~121102.\footnote{For sources with low redshift, the
comoving distance $\propto z$, thus the comoving volume $\propto z^3$.} Even in
such a case, the inclusion of the measured redshift does not provide a worse
constraint compared with the case where we are uninformative about the
redshift. Were the measured redshift larger, the constraint would be better.
For now, the constraint on the photon mass is limited by our
assumptions about the uncertainties from the Milky Way, the IGM, and the host
galaxies of FRBs (cf.  Section~\ref{sec:bayes}). Better determination of these
contributions leads to tighter limits for individual FRBs, and when combined
through Eq.~(\ref{eq:bayes}), a better combined constraint results. We expect
all three points listed above will make progresses in observations soon.

The reason that in our analysis those FRBs without the redshift measurement
still contribute to the constraint is the use of an uninformative prior for the
redshift and the inclusion of it in a Bayesian way. In addition to the
uniform-in-volume prior used in Section~\ref{sec:res}, we here use another
physical prior that traces the star formation rate for a robustness test.  We
use the fit for the star formation rate given by~\citet{Yuksel:2008cu},
\begin{eqnarray}\label{eq:sfr}
  \dot \rho_* (z) &=& \dot \rho_0 \left[ \left( 1+z \right)^{a\eta} + \left(
  \frac{1+z}{B} \right)^{b\eta} + \left( \frac{1+z}{C} \right)^{c\eta}
\right]^{1/\eta} \,,
\end{eqnarray}
where $\dot \rho_0=0.02 \, M_\odot \, {\rm yr}^{-1} \, {\rm Mpc}^{-3}$,
$a=3.4$, $b=-0.3$, $c=-3.5$, $\eta=-10$, $B=\left( 1+z_1 \right)^{1-a/b}$,
$C=\left( 1+z_1 \right)^{(b-a)/c} \left( 1+z_2 \right)^{1-b/c}$ with the
breaking points $z_1=1$ and $z_2=4$.  We obtain a combined limit $7.5 \times
10^{-51} \, {\rm kg}$ ($1.3 \times 10^{-50} \, {\rm kg}$) at 68\% (95\%)
confidence level, showing that reasonable changes in the prior of redshift do
not lead to large difference. The slight improvement here results from the fact
that the star formation rate in Eq.~(\ref{eq:sfr}) favours larger $z$ when $z
\leq 1$ where most FRBs in Table~\ref{tab:frbcat} reside.

Lastly, worthy to mention that, because FRBs are distributed nearly
isotropically in the sky (see Figure~\ref{fig:frbsky}), they will also be
useful to constrain the anomalous anisotropic inertial mass tensor of
photons~\cite{Lammerzahl:1998qp, Kostelecky:2002hh, Kostelecky:2009zp} in a
similar way that pulsars are used to set constraints on a Lorentz-violating
tensor~\cite{Shao:2014oha}. We leave this point for future work.


%

\end{document}